\documentclass[smallextended]{svjour3}       
\smartqed  
\usepackage{graphicx}
\usepackage{natbib}
\usepackage{amsmath}
\usepackage{url}

\usepackage{mathptmx}      
\begin{document}

   \title{The Solar Line Emission Dopplerometer project}

   \titlerunning{The Solar Line Emission Dopplerometer (SLED)}

   \authorrunning{J.-M. Malherbe et al.}

   \author{Jean-Marie Malherbe
          \and
          Pierre Mein
          \and
          Fr\'{e}d\'{e}ric Say\`{e}de
           \and
          Pawel Rudawy
           \and
          Kenneth Phillips
           \and
          Francis Keenan
           \and
          Jan Ryb\'{a}k
          }

   \institute{J.-M. Malherbe \at
              LESIA, Observatoire de Paris, PSL Research University, CNRS, 92195 Meudon,
              France\\
              \email{jean-marie.malherbe@obspm.fr}
         \and
             P. Mein \at
             LESIA, Observatoire de Paris, PSL Research University, CNRS, 92195 Meudon,
             France\\
             \email{pierre.mein@club-internet.fr}
         \and
              F. Say\`{e}de \at
              GEPI, Observatoire de Paris, PSL Research University, CNRS, 92195 Meudon,
              France\\
              \email{frederic.sayede@obspm.fr}
         \and
              P. Rudawy \at
              Astronomical Institute, University of Wroc{\l}aw,
              Poland\\
              \email{rudawy@astro.uni.wroc.pl}
         \and
              K.J.H. Phillips \at
              Earth Sciences Department, Natural History Museum, London SW75BD, United
              Kingdom\\
              \email{kennethjhphillips@yahoo.com}
         \and
              F.P. Keenan \at
              Astrophysics Research Centre, School of Mathematics and Physics, Queen's University Belfast, United
              Kingdom\\
              \email{f.keenan@qub.ac.uk}
         \and
              J. Ryb\'{a}k \at
              Astronomical Institute, Slovak Academy of Sciences, 05960 Tatransk\'{a} Lomnica,
              Slovakia\\
              \email{rybak@astro.sk}
              }

\date{Received: 2 April 2021 / Accepted: date}

\maketitle

\begin{abstract}
Observations of the dynamics of solar coronal structures are
necessary to investigate space weather phenomena and global heating
of the corona. The profiles of high temperature lines emitted by the
hot plasma are usually integrated by narrow band filters or recorded
by classical spectroscopy. We present in this paper details of a new
transportable instrument (under construction) for imaging
spectroscopy: the Solar Line Emission Dopplerometer (SLED). It uses
the Multi-channel Subtractive Double Pass (MSDP) technique, which
combines the advantages of both filters and narrow slit
spectrographs, i.e. high temporal, spatial and spectral resolutions.
The SLED will measure at high cadence (1 Hz) the line-of-sight
velocities (Doppler shifts) of hot coronal loops, in the forbidden
lines of FeX 6374 \AA~ and FeXIV 5303 \AA. It will follow the
dynamics of fast evolving events of solar activity such as flares or
Coronal Mass Ejections (CMEs), and also study coronal heating by
short period waves. Observations will be performed with the
coronagraph at the Lomnick\'{y} \v{S}t\'{i}t Observatory (LSO, in
Slovakia) or during total eclipses. The SLED will also observe the
dynamics of solar prominences in H$\alpha$ 6563 \AA~ or He D3 5876
\AA~ lines when mounted on the Bia{\l}k\'{o}w coronagraph (near
Wroc{\l}aw, Poland). It is fully compatible with polarimetric
measurements by various techniques.

\end{abstract}


\section{Introduction} \label{sec:Intro}

Solar activity is the primary driver of space weather
\citep[e.g.,][]{Schrijver2012, Pomoell2018}. It occurs over
timescales ranging from seconds to several minutes in solar flares
and CMEs \citep{Shibata2011}, which are fast evolving events and
highly dynamic phenomena, originating in the solar atmosphere from
non-potential energy stored in magnetic fields. Observational data
for coronal loops, and their main properties and models, are
reviewed by \cite{Reale2010}.

The hot corona has been imaged in the extreme ultraviolet (EUV) at
45 s cadence by the Atmospheric Imaging Assembly onboard Solar
Dynamics Observatory (SDO/AIA) since 2010 in many wavebands
containing highly ionized lines of iron (171, 193, 211, 335, 94, 131
\AA~ respectively for temperatures of 0.6, 1.0, 2.0, 2.5, 6.0 and
10.0 MK). STEREO, since 2006, uses similar wavebands (171, 195, 284
\AA) plus two coronagraphs. The Solar and Heliospheric Observatory
(SOHO), launched in 1995, is still active in these wavebands (EIT
telescope) and with the C2/C3 wide field and white light
coronagraphs of LASCO. Solar Orbiter, launched in 2020, will soon
offer, with the EUI instrument \citep{Rochus2020}, coronal images in
wavebands at 171 and 335 \AA. In parallel, high altitude
coronagraphs allow observations at much higher cadence (sub-second),
such as the SECIS instrument (\cite{Phillips2000},
\cite{Ambroz2010}) working in white light and in the green coronal
line of FeXIV 5303 \AA.

The corona is also observed during total solar eclipses in several
high temperature emission lines including FeX 6374 \AA~ (red line)
and FeXIV 5303 \AA~ (green line), as well as in infrared (IR) lines
such as FeXI 7892 \AA~ (\cite{Habbal2011}, \cite{Boe2018},
\cite{Rudawy2019}).

However, none of the above imaging instruments can measure the
plasma velocities in the corona. Various instrumental techniques,
such as radar echoes, have been used in the past to probe the solar
corona and measure Doppler shifts. For example, \cite{Chisholm1964}
found frequency shifts in 38.25 MHz radar signals reflected by the
corona. \cite{James1970} suggested that such echoes could be due to
compressional waves related to coronal heating or to mass motions
with velocities less than 50 km s$^{-1}$. \cite{Desai1982} carried
out Fabry-P\'{e}rot interferometric observations of the coronal
FeXIV green line and confirmed mass motions in the 30-50 km s$^{-1}$
range. LASCO/C1, onboard SOHO, was operational from 1995 to 1998 and
able to scan the green (FeXIV) and red (FeX) lines with a
Fabry-P\'{e}rot tunable filter (respectively 0.65 and 0.85 \AA~
spectral resolution, \cite{Brueckner1995}). Mierla (PhD Thesis,
G\"{o}ttingen, 2005) and \cite{Mierla2008} reported velocities of
the order of 10 km s$^{-1}$ in the quiet corona, but C1 was not able
to measure velocities for short duration dynamic events such as
CMEs, due to the long time needed (15 minutes) to scan the line
profiles. Among ground-based coronagraphs, the Coronal Multi-channel
Polarimeter (CoMP, \cite{Tomczyk2008}) is a tunable birefringent
filter and full Stokes polarimeter working in the IR lines of FeXIII
10747 \AA~ and 10748 \AA~ (1.3 \AA~ FWHM). CoMP is able to measure
line-of-sight (LOS) velocities and magnetic fields, with temporal
resolution of about 15 s and spatial scale of 4.35$''$.
\cite{Morton2016} used CoMP to identify transverse-like waves that
peaked at frequencies of around 3 mHz.

In spectroscopy, multi-slit spectra of the green line have been
reported by \cite{Livingston1980} and \cite{Livingston1982}.
\cite{Singh2002} have investigated the properties of the coronal
FeXIV and FeX emission lines, using slit spectroscopy at high
spectral resolution (respectively 0.03 and 0.06 \AA). Spectra of the
FeXIV line (0.02 \AA~ resolution) were also recorded by
\cite{Sakurai2002}. \cite{Minarovjech2003} reported the detection of
oscillations in the green line observed at LSO and at Norikura
(nearly simultaneously) using slit spectroscopy. \cite{Lee2021}
combined simultaneous Doppler measurements from CoMP (FeXIII 10747
A) and EIS onboard Hinode (FeXII 195 \AA, FeXIII 202 \AA) and found
a good correlation above active regions.

The SLED instrument described in this paper will provide an unique
opportunity to measure plasma velocities in the coronal green and
red lines at high cadence (1 Hz), by combining the advantages of
both spectroscopy and tunable filters. Our paper is organized as
follows. Section ~\ref{sec:MSDP} summarizes the principles and
history of the imaging spectroscopy concept used by the SLED in a
state-of-the-art manner, while Section ~\ref{sec:SLED} presents the
main scientific goals of the SLED. The capabilities of the new
instrument are described in Section ~\ref{sec:SLEDcap}, while
Section ~\ref{sec:design} provides details concerning the optical
design.

\section{The Multichannel Subtractive Double Pass (MSDP)} \label{sec:MSDP}

\begin{figure}
   \centering
   \includegraphics[width=\textwidth]{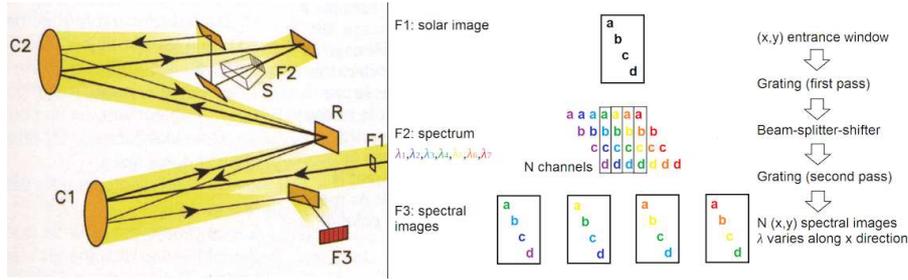}
      \caption{The principle of the MSDP spectroscopy.
   Left: the spectrograph; F1 = entrance window; C1 = collimator; C2 = chamber objective;
   R = grating; F2 = spectrum; S = slicer (beam-splitter, beam-shifter); F3 = multichannel spectra image and camera.
   Right: the three successive steps of a schematic MSDP with N = 4 channels; top = solar image at entrance window F1; middle =
   slicer S in the spectrum at F2 (first pass, the colours of the letters correspond
   to the light
   dispersion); bottom: N = 4 spectra images at F3 (after second subtractive pass), with varying wavelength
   along the x-direction).}
         \label{MSDP}
   \end{figure}

The SLED is a promising, highly optimized and compact version of the
MSDP spectrograph; the principle behind the instrument is
illustrated in Figure~\ref{MSDP}. The MSDP was first described by
\cite{TOUR77}, and has been upgraded many times over the last four
decades on several telescopes in Europe (Meudon Solar Tower, Pic du
Midi Turret Dome, Vacuum Tower Telescope and THEMIS at Tenerife,
Bia{\l}k\'{o}w coronagraph; see \cite{Mein2021}).

It is an imaging double pass spectrograph using a 2D rectangular
entrance window (F1) and a slicer (S). In the spectrum (F2), after a
first pass on the grating (R), the slicer has two functions: it
selects N channels (beam-splitting) and realigns the N channels
(beam-shifting) before the subtraction of the dispersion by the
second pass on the grating. The beam-splitter can be either slit or
micro-mirror based, while the beam-shifter uses either prisms or
mirrors. Output of the MSDP is composed of N contiguous
spectra-images (N increased from 7 to 24 during four decades of
advances). There is a constant wavelength step between each
spectra-image, but inside each, the wavelength varies linearly along
the x-direction. This technique allows the derivation of data cubes
(x, y, $\lambda$), where the three coordinates are simultaneous
while extracted from a single exposure. Hence, the MSDP combines the
advantages of imaging filters and spectroscopy, and the spatial
resolution is identical to that provided by filters. The FOV depends
on the focal length of the telescope and the size of the window
(F1); the size can be as large as 60$''$ in the x-direction and
500$''$ in the y-direction. Spectral resolution is in the range 0.03
- 0.3 \AA~ (according to the focal length of the spectrograph and
the slicer technology), while the cadence may reach several
frames/s, depending on the photon flux.

\begin{figure}
   \centering
   \includegraphics[width=\textwidth]{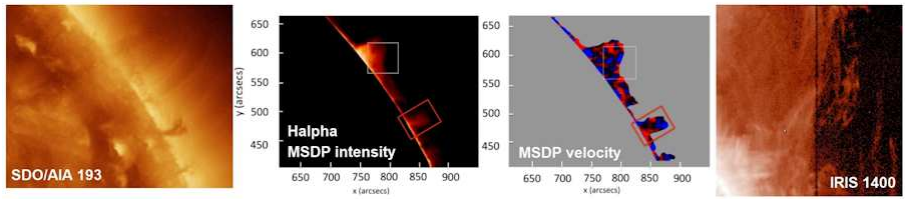}
      \caption{Typical coordinated observations (24 September 2013 at 12:22 UT) with SDO/AIA (193 \AA, left),
      the Meudon MSDP (H$\alpha$ intensity and Doppler velocity) and IRIS (slit jaw at 1400 \AA, right). The white rectangle (prominence)
      delineates the IRIS FOV
       (50 $''$ $\times$ 50 $''$), while the red box
      indicates the tornado.}
         \label{dop}
   \end{figure}

Figure~\ref{dop} shows typical processed observational data of
prominences and tornadoes which have been performed with the MSDP in
coordination with space-borne instruments such as IRIS
\citep{SCHMIEDER2014}.

\begin{figure}
   \centering
   \includegraphics[width=\textwidth]{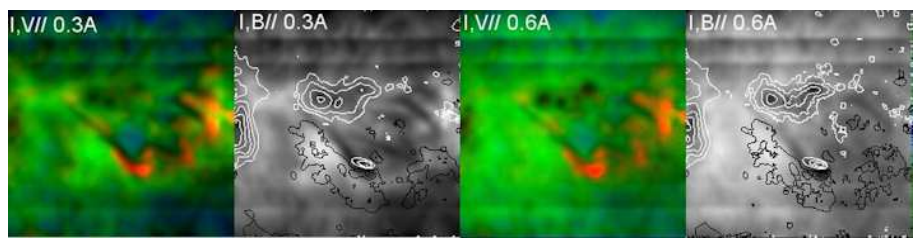}
      \caption{Polarimetric test on an active region with the Meudon MSDP.
      Intensity I, LOS velocity V//, and line of sight magnetic field B//
       at two positions in the H$\alpha$ line (0.3 \AA~ and 0.6 \AA~ chords above line centre) corresponding
      to two different altitudes. B// is represented by contours (white/black for North/South
      polarities). V// is colour-coded (blue/red shifts). FOV 300$''$ x 300 $''$.}
         \label{IVB}
   \end{figure}

   The MSDP is fully compatible with polarimetric observations. Many
   techniques have been used on various telescopes: Stokes V with
   liquid crystals at Meudon (Figure~\ref{IVB}) and Pic du Midi
   \citep{Roudier2006}; static birefringent plates at THEMIS for
   Stokes Q, U, V. The polarimeter is installed in the image plane near
   the entrance window F1 of Figure~\ref{MSDP}.

\section{The scientific objectives of the SLED} \label{sec:SLED}

The SLED has two main goals which can be achieved owing to its
capability to produce Dopplergrams for the whole FOV at high cadence
(1 Hz).

\subsection{Highly dynamic events}

SLED will observe the dynamics of coronal loops in active regions,
flares and CMEs, driving solar-terrestrial interactions and space
weather events. Crucially, it will provide a new diagnostic via
plasma velocity measurements (Doppler shifts) inside hot loops at
high temporal resolution. This outstanding feature can be combined
with data obtained by AIA onboard SDO (intensity of several lines
between 0.6 MK and 10 MK at lower cadence). SLED, with its LOS
velocity measurements, will impose new constraints on models of hot
coronal loops and their time evolution. While line intensity
information allows us to visualize qualitatively plasma motions in
the plane of the sky, Doppler shifts provide a direct determination
of the LOS velocity. SLED will be operational well before the
forthcoming solar maximum of cycle 25 (predicted in 2025) and will
measure velocities for highly dynamic events which play a major role
in space weather.

\subsection{Coronal heating}

The SLED will search for high-frequency, wave-like variations in the
plasma velocities over small spatial scales in the solar corona,
particularly in loops associated with active regions. It allows
quantitative studies of wave processes contributing to the plasma
heating, energy balance of the corona and tests of coronal heating
models. SLED will not search for intensity fluctuations but for
motions derived from Doppler shifts in the green and red forbidden
lines, as seen by \cite{Tomczyk2007} in the near-infrared FeXIII
lines.

The physical process involved in the heating of the solar corona is
not yet definitively identified \citep{Klimchuk2006}. The main
mechanisms that have been studied fall into two categories: heating
by numerous small magnetic reconnection events (nano-flares) or by
the dissipation of Alfv\'{e}n or magnetohydrodynamic (MHD) waves in
coronal loops.

Observations with CoMP at Sacramento Peak \citep{Tomczyk2007} have
shown evidence for Alfv\'{e}n-type wave motions around active
regions at the limb. The waves were upward-propagating with low
frequencies (3.5 mHz). However, the total energy flux of these waves
is insufficient for the coronal heating. The investigation of much
higher-frequency waves (1 Hz) is not possible with current space
missions owing to telemetry limitations, and therefore has been
undertaken during total solar eclipses with ground-based
instruments. In these studies, the corona has been imaged using
narrow-band filters (typically up to 5 \AA) in the FeXIV green line
or FeX red line with fast CCD cameras. Several observational
campaigns were organized with the SECIS instrument
(\cite{Phillips2000}, \cite{Rudawy2004}, \cite{Rudawy2010},
\cite{Rudawy2019}), but a careful analysis showed that the short
period intensity fluctuations are questionable in terms of waves.
For this reason, the SLED will search instead for signatures of
Doppler shifts in wave-like phenomena.

\section{The SLED capabilities} \label{sec:SLEDcap}

The core of the SLED is the 24 channels slicer. It is designed for
F/30 beams and can be easily coupled to various existing telescopes.
Hence, it is envisaged to become a permanent instrument for the
large Bia{\l}k\'{o}w coronagraph (prominence observations) or for
the high-altitude LSO coronagraph (high temperature emission lines).
It will also be used for total eclipse campaigns, as it is a
portable spectrograph. In summary, the SLED can be mounted on the
following telescopes with easy focal length adaptation:

\begin{itemize}
  \item the 2.0 m focal length / 0.20 m aperture Celestron telescope at Wroc{\l}aw (6.0 m
  equivalent focal
  length with Barlow 3.0 $\times$)
  and heliostat (or equatorial mount) for total eclipse campaigns, providing a 150$''$ $\times$ 1000$''$ FOV with 2.1$''$ pixel
  sampling.
  \item the 4.0 m focal length / 0.2 m aperture Zeiss coronagraph \citep{Lexa1963} at LSO (6.0 m
  equivalent focal
  length with Barlow 1.5 $\times$), for coronal observations (hot plasma), 150$''$ $\times$
  1000$''$ FOV with 2.1$''$ pixel sampling.
  \item the 14.5 m focal length coronagraph (0.50 m aperture) at
  Bia{\l}k\'{o}w observatory
  providing a 62.5$''$ $\times$ 450$''$ FOV with 0.9$''$ pixel sampling (possibly two times larger with
  0.25 m diaphragm and focal length reduced to 7.25 m), for prominence observations (cold
  plasma).
\end{itemize}

Coronal lines will be observed at the LSO coronagraph or during
total eclipses (the Barlow is the only element that will change).
However, SLED can also observe in H$\alpha$ 6563 \AA~ or the He D3
5876 \AA~ lines to study the dynamics of prominences with the
Bia{\l}k\'{o}w coronagraph. Indeed, the SLED would considerably
improve the capabilities of the MSDP already operating there with
only 9 channels.

\begin{figure}
   \centering
   \includegraphics[width=\textwidth]{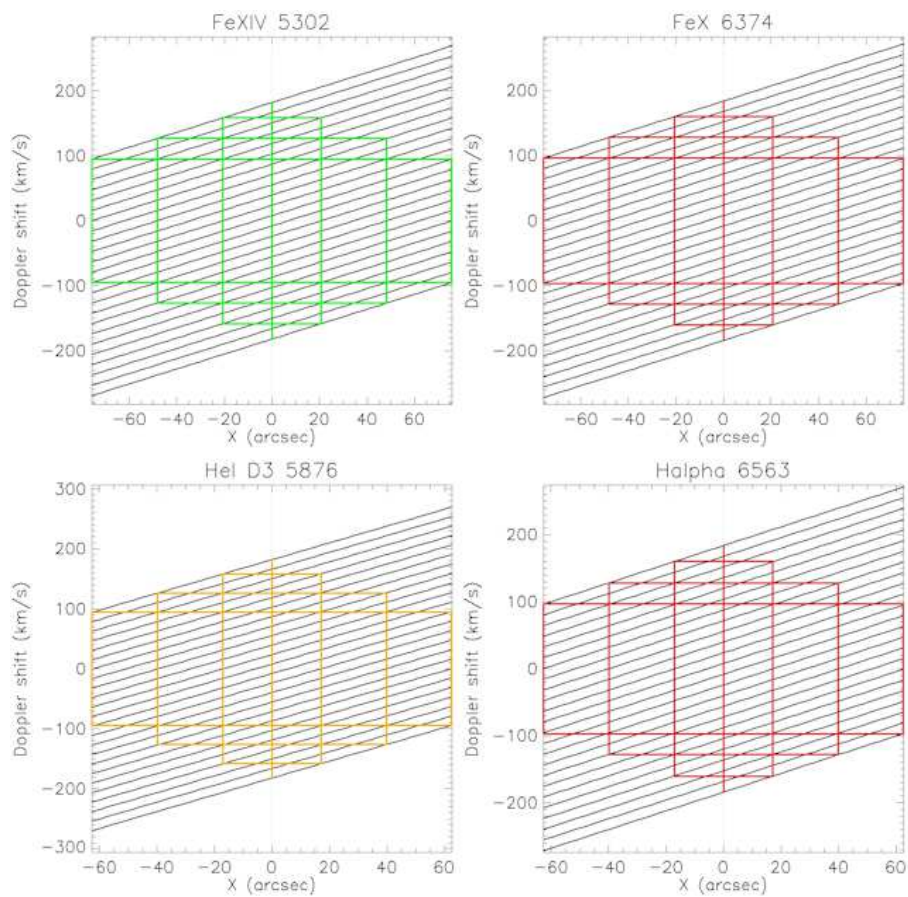}
      \caption{Wavelength functions of the 24 channels of the SLED.
      Abscissae: the x-direction ($''$); ordinates: the wavelength variation
      converted to velocities (km s$^{-1}$). Top:
      coronal lines (left: FeXIV, right: FeX) for LSO coronagraph or eclipses; bottom: prominence lines (left: He D3,
      right: H$\alpha$) for Bia{\l}k\'{o}w coronagraph.}
         \label{bandpass}
   \end{figure}

The wavelength transmissions of the 24 channels in terms of Doppler
shifts (converted to velocities, positive for blue shifts) are
displayed in Figure~\ref{bandpass} for the coronal green and red
lines (LSO and eclipse telescopes, 6.0 m equivalent focal length)
and for the prominence lines (Bia{\l}k\'{o}w coronagraph, 7.25 m
focal length). Note that the channels are not monochromatic; the
wavelength varies linearly in the x-direction (along the width of
the FOV). Figure~\ref{bandpass} shows that large velocities can be
determined everywhere in the full FOV. However, the three
rectangular selections (in terms of x-direction and wavelength) show
that there is a linear relationship between the highest measurable
velocities and the FOV width. For the green line, the maximum
Doppler shift (converted to velocity) is given by the relation
$\lvert v_{max} \rvert = -1.17 \lvert x \rvert + 183$ km s$^{-1}$
where x is the abscissa ($''$) along the 2D FOV. Hence, the SLED is
well adapted to the highly dynamic phenomena and the fast evolving
events associated with solar activity.

\begin{figure}
   \centering
   \includegraphics[width=\textwidth]{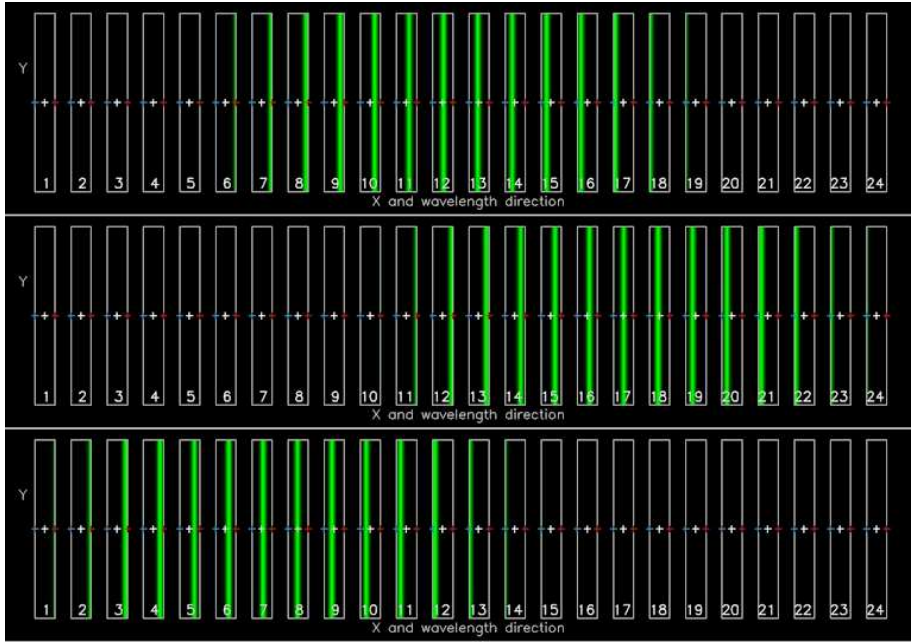}
      \caption{A simulation of spectra images of the FeXIV green
      line (no velocity, top) and
      with redshifts (-75 km s$^{-1}$, middle) or blueshifts (+75 km s$^{-1}$, bottom).
      The crosses refer to the three specific locations in the x-direction (left, centre, right of the FOV)
      indicated in Figure~\ref{lines}.}
         \label{green}
   \end{figure}

\begin{figure}
   \centering
   \includegraphics[width=\textwidth]{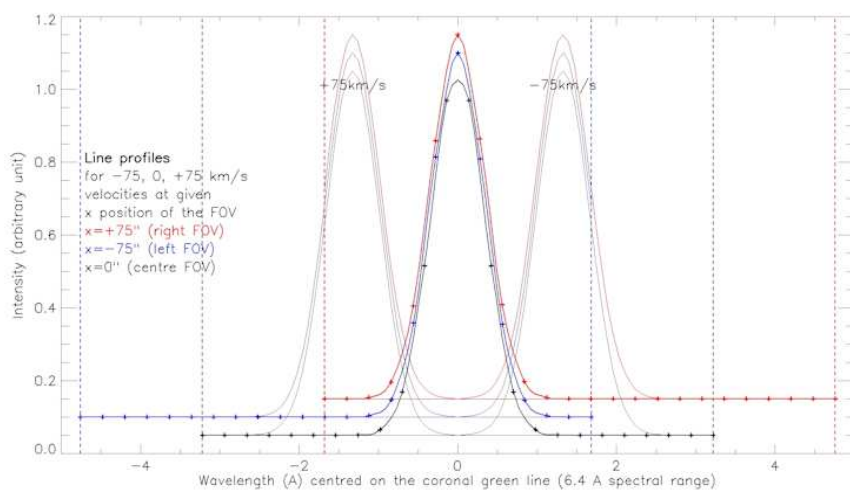}
      \caption{Line profiles at the three locations selected in
      Figure~\ref{green},
      with LOS velocities of 0, -75 and +75 km s$^{-1}$.
      Black: profiles at the centre of the FOV. Blue/red: profiles
      respectively at the left/right sides of the FOV. Crosses show
      the measurement points; a cubic interpolation is applied between them. The
      spectral range is 6.5 \AA; it shifts from the left to the
      right sides
      of the FOV, so that the centering of the line varies in x-direction.}
         \label{lines}
   \end{figure}

Figure~\ref{green} shows a simulation of the 24 channels for the
FeXIV green line, the profile of which is assumed to be Gaussian.
The FWHM of the profile is 0.8 \AA~ \citep{Singh2002} and the line
is centred between channels 12-13 (top row). The middle and bottom
parts of Figure~\ref{green} show how the line is affected by LOS
velocities of respectively -75 km s$^{-1}$ (red shift) and +75 km
s$^{-1}$ (blue shift). Three points indicated by crosses (left,
centre and right of the FOV in the x-direction) refer to the
profiles of Figure~\ref{lines}. In the centre of the FOV, the
spectral range provided by the 24 channels is [-3.25 \AA, +3.25
\AA], allowing the measurement of Doppler shifts in the range [-2.85
\AA, +2.85 \AA] (to take into account the line width), corresponding
to velocities up to 160 km s$^{-1}$. Towards the edges, the spectral
domain has the same width (6.5 \AA) but is shifted: [-4.8 \AA, +1.7
\AA] or [-1.7 \AA, +4.8 \AA] respectively for the left and right
sides; there velocities up to 75 km s$^{-1}$ can still be
determined. Steady flows in coronal loops exhibit velocities which
are often smaller, but in the cases of ejected material, the SLED
will be able to measure velocities not far from the sound speed (150
km s$^{-1}$).

\begin{figure}
   \centering
   \includegraphics[width=\textwidth]{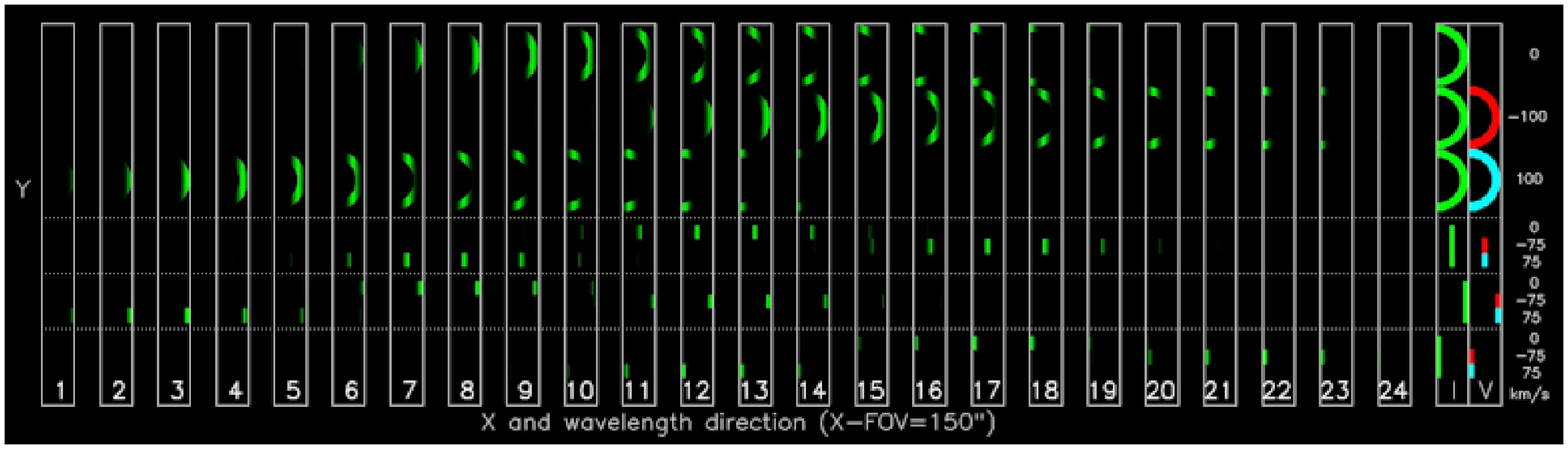}
      \caption{Simulation of the 24 channels of the SLED for a model
      of observations shown at right in the columns I (intensity) and V (velocity).}
         \label{simu}
   \end{figure}

Figure~\ref{simu} shows another simulation of the 24 channels of the
SLED corresponding to several models of structures displayed in the
column I (intensity) and V (LOS velocity in km s$^{-1}$), under the
assumption of Gaussian line profiles (0.8 \AA~ FWHM). In practice,
the 24 channels are approximately parallelograms detected by the
data processing algorithm and transformed into 24 rectangles of the
same size, after correcting for geometrical distortions.
Subsequently, the algorithm uses the information from the 24
channels of the spectra images to recover the line profiles and
derive intensity and velocity maps, after correction by the flat
field (observed generally at disk centre). The restoration of the
line profiles is based on the transmission curves of
Figure~\ref{bandpass} and cubic interpolation between sampling
points (0.28 \AA~ step). As standard polynomial interpolations
between them (0.07 \AA~ step) generate errors, in particular in the
line core, we propose a two-step method to restore as best as
possible the profiles.

\begin{figure}
   \centering
   \includegraphics[width=\textwidth]{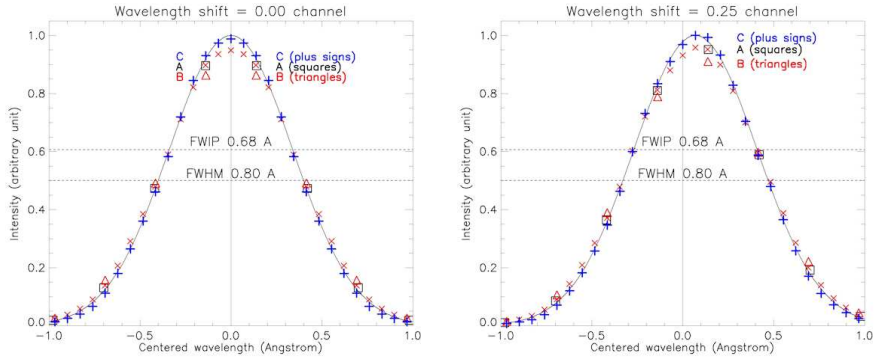}
      \caption{Improved interpolation method proposed for the SLED.
      Left and right pannels just differ by the distribution of sampling points
      (which depend on the pixel location in the x-direction and/or the Doppler
      shift; the right example has a shift of 1/4 channel = 0.07 \AA).
      Solid (black) line: theoretical Gaussian
      profile.
      Black squares (A): sampling points by the slicer (0.28 \AA~ step and bandwidth). Red crosses: cubic
      interpolation between sampling points ("first order" restored profile).
      Red triangles (B): sampling points of the "first order" profile (0.28 \AA~ step and bandwidth).
      Blue signs (C): corrected positions
        (B $\rightarrow$ C, such that BC = 2 BA) providing the "second order" restored line profile.
      FWIP = Full Width at Inflexion Points, FWHM = Full Width at Half Maximum. }
         \label{interpol}
   \end{figure}

In Figure~\ref{interpol}, we have plotted a theoretical Gaussian
profile in black. This is then sampled by the slicer of the SLED
(0.28 \AA~ for both step and bandwidth) and the measurement points
are indicated by black squares (A). The cubic interpolation between
points A provides the red crosses (0.07 \AA~ step). This
"first-order" restored profile reveals errors, mainly in the
vicinity of the line core. In order to compensate for these, we
compute a "second-order" profile (blue signs, C points). For that
purpose, the "first-order" profile is re-sampled similarly (0.28
\AA~ step and bandwidth) providing the red triangles (B points).
Then, we apply a correction such that AC = BA (thus BC = 2 BA),
providing the "second-order" restored profile indicated by the blue
signs (C). This two-step method provides restored profiles (the C
points) which are the most accurate above and below the inflexion
points (respectively in the line core and wings).

\begin{figure}
   \centering
   \includegraphics[width=\textwidth]{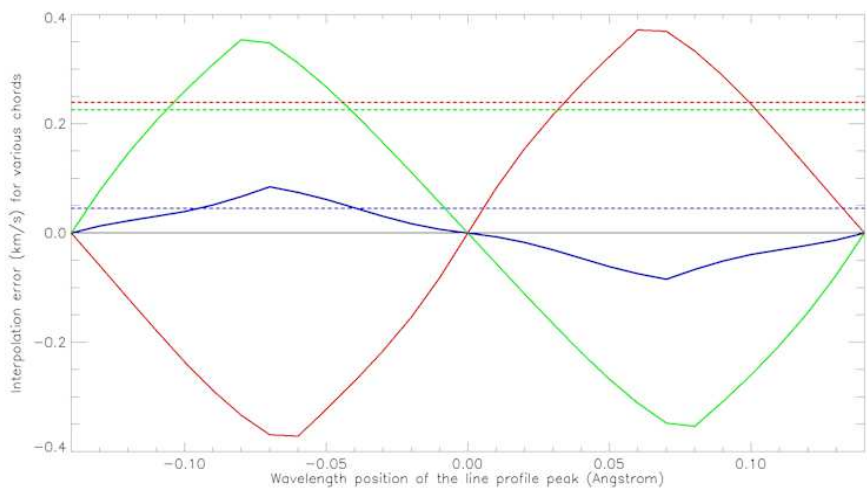}
      \caption{Velocity error for different chords as a function of
      the wavelength sampling of the line profile. The wavelength range (1 channel = 0.28 \AA)
      is the period of the plot.
      For positions -0.14, 0 , +0.14 \AA, sampling
      points are symmetrical and the errors vanish. The chord widths are 0.70, 0.84 and 0.56 \AA, respectively
      for blue, green and red solid lines. The corresponding RMS error levels (0.05,
      0.22, 0.24 km s$^{-1}$ respectively) are indicated by the dashed lines. }
         \label{chords}
   \end{figure}

LOS velocities are derived from the restored profiles using the
bisector technique. For that purpose, we choose a chord of given
width and compute the wavelength position of the middle of the
chord. The simulation allows us to assess the precision of velocity
measurements for different chords. Figure~\ref{chords} shows that
the best precision is always obtained with a chord of 0.70 \AA~
between the inflexion points (because the slope is maximum there).
For the FeXIV green line, assuming a gaussian shape, we found a FOV
averaged error of $\pm$ 0.05 km s$^{-1}$; locally, it may reach
$\pm$ 0.08 km s$^{-1}$ (depending on the location of the sampling
points along profiles). Below or above the inflexion points, errors
are higher ($\pm$ 0.23 km s$^{-1}$).

\begin{figure}
   \centering
   \includegraphics[width=\textwidth]{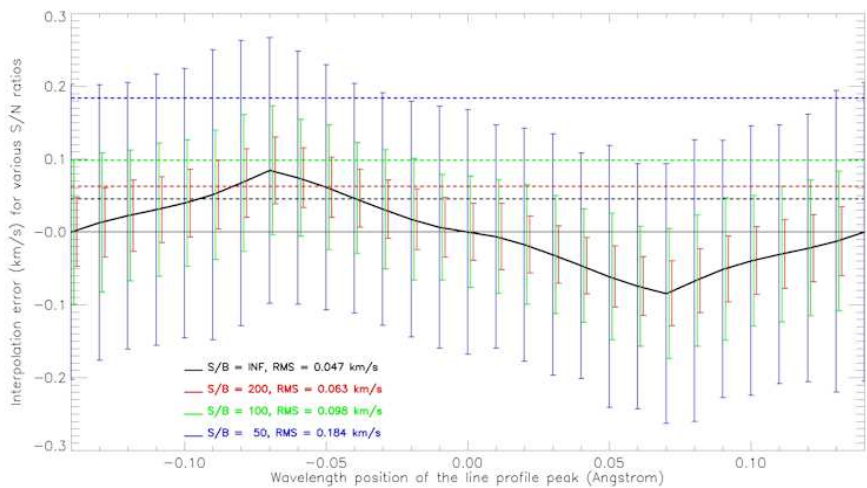}
      \caption{Influence of photon noise on the precision of
      velocity measurements as a function of
      the wavelength sampling of the line profile. The wavelength range (1 channel = 0.28 \AA)
      is the period of the plot.
      The error bars indicate the effect of the photon noise for S/N = 50, 100, 200
      (respectively blue, green and red bars). The corresponding RMS error levels (0.18,
      0.10, 0.06 km s$^{-1}$ respectively) are indicated by the dashed lines.}
         \label{noise}
   \end{figure}

It is well known from the signal-to-noise ratio (S/N) of
observations that photon noise is an important source of error. In
order to simulate this, we introduced a Poisson-type noise on line
profiles and found a FOV averaged error of $\pm$
  0.35, $\pm$ 0.18, $\pm$ 0.10, $\pm$ 0.06 km
  s$^{-1}$ for respectively S/N = 25, 50, 100, 200
  (Figure~\ref{noise}), in the case of velocity measurements at the
  inflexion points of the line. The sCMOS detector cannot exceed a S/N ratio of 150 for a single
exposure (30000 electrons full well capacity), providing a typical
velocity error of $\pm$ 0.07 km s$^{-1}$. Photon noise is negligible
for S/N ratios above 200. For 1 Hz observing cadence we expect for
the SLED the S/N ratio of 50, so that the sampling and cubic
interpolation errors ($\pm$ 0.05 km s$^{-1}$) will be dominated by
the photon noise ($\pm$ 0.18 km s$^{-1}$). At slower cadences (0.25
Hz or less), the restoration will just be limited by the
interpolations.

The movie (see appendix) presents a simulation of the SLED channels
for the coronal green and red lines emitted by artificial structures
with varying velocities in the interval [-100 km s$^{-1}$, +100 km
s$^{-1}$], together with the associated line profiles.

As a future option, the SLED could operate in polarimetric mode for
magnetic field strength and direction determination. \cite{Mein2021}
describe a full Stokes polarimeter optimized for the MSDP imaging
spectroscopy. It uses a calcite beam splitter which reduces the FOV
in the x-direction by a factor two, and hence provides two
simultaneous measurements (I+S, I-S with S = Stokes Q, U, V in
sequence). In order to preserve the FOV in the x-direction, an
alternative method is to measure sequentially I+S and I-S, as
implemented by \cite{Roudier2006} at the MSDP of the Pic du Midi
turret dome. The polarimeter takes place at the entrance window of
the spectrograph and is based on Liquid Crystal Variable Retarders
(LCVR) allowing fast observations. While one crystal is needed for
Stokes Q and V, two are required for U. The signals delivered by the
optional polarimeter would be:

$ \frac{1}{2} [ I \pm (Q \cos\delta_{2} + \sin\delta_{2} (U
\sin\delta_{1} - V \cos\delta_{1}) )]$

\noindent where $\delta_{1}$ and $\delta_{2}$ are the two LCVR
retardances.

\section{The SLED optical design} \label{sec:design}

The spectral resolution (0.28 - 0.34 \AA) and coverage (6.5 \AA) of
the SLED is well adapted to coronal line widths (typically 0.8 \AA)
and large Doppler shifts ($\pm$ 100 km s$^{-1}$ $\approx$ $\pm$ 2.0
\AA), and its large FOV (150$''$ $\times$ 1000$''$, 2.1$''$ spatial
sampling) is compatible with the size of coronal structures
($10^{5}$ km or more). We extrapolated from SECIS observations that
the expected cadence (only limited by the photon flux of the 0.20 m
aperture) is one frame/s for a S/N ratio of 50, and several frames/s
if a smaller S/N ratio is acceptable (depending on the scientific
program). The slicer is built by Paris observatory; the spectrograph
design, prepared in Paris, will be assembled and tested by
Wroc{\l}aw to be fully operational well in advance of the next solar
maximum in 2025. SLED will also be available for the total solar
eclipse of 8 April 2024, visible from North America.

\begin{figure}
   \centering
   \includegraphics[width=\textwidth]{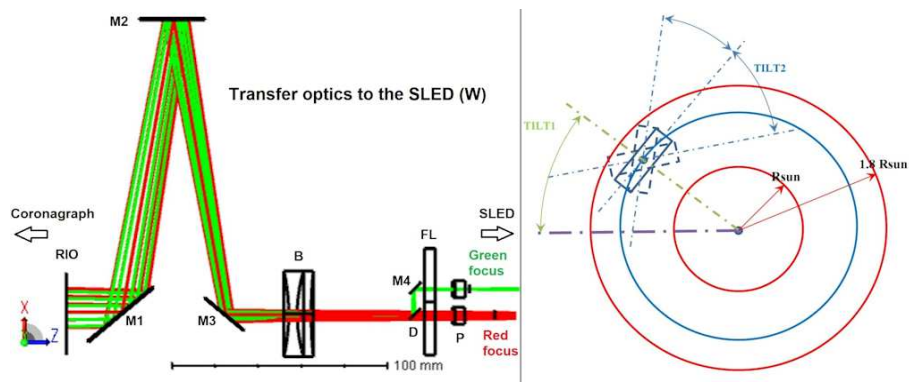}
      \caption{Left: transfer optics from the telescope to the SLED spectrograph
      in the case of
      the LSO Zeiss telescope. RIO = re-imaging objective (4 lenses) of the coronagraph (4.0 m secondary focus at F/20); (M1, M2,
      M3) = field rotator; B = Barlow (diverging doublet providing 6.0 m equivalent focus at F/30); D = dichroic
      mirror; FL = two field lenses; P = two compensating plates. Right: the positioning system
      of the 2D entrance windows; TILT1 corresponds to the axial rotation of the Zeiss mount; TILT2 comes from the rotator, as
      well as the adjustable distance from the limb (1.0 to 1.8 solar radii).}
         \label{design}
   \end{figure}

The transfer optics to the SLED is detailed by Figure~\ref{design},
for the case of the LSO coronagraph (0.2 m diameter, 3.0 m primary
focus, 4.0 m secondary focus). The post-focus instrument mechanical
interface allows the inspection of the entire solar limb (TILT1
angle). SLED includes a field rotator (M1, M2, M3 mirrors) allowing
us to explore the solar corona up to 0.8 solar radii above the limb.
The 2D entrance window will be tangential to the limb, but can also
be inclined (TILT2 angle) to fit better to the coronal loop's
topology. The diverging doublet B provides a F/30 focus (6.0 m
equivalent focal length). Red and green light beams are separated by
the dichroic mirror D to form two solar images at the focus of the
SLED spectrograph. In Figure~\ref{design}, P are compensating plates
(for accurate focusing of green and red lines) and FL are field
lenses to reject the entrance pupil at large distances.

\begin{figure}
   \centering
   \includegraphics[width=\textwidth]{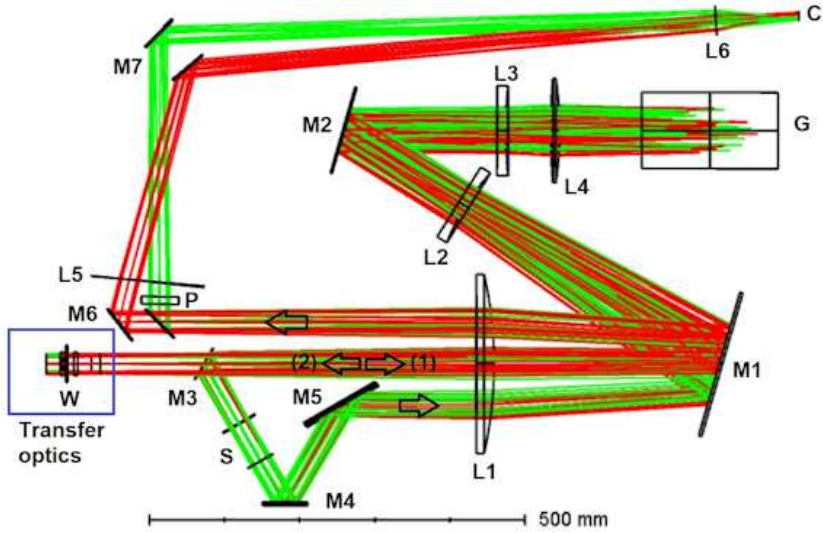}
      \caption{The SLED optical design;
  W = transfer optics containing entrance windows (see details in Figure~\ref{design});
 L = lenses or objectives; M = flat folding
 mirrors; G = grating; S = slicer;
 C = sCMOS camera for two simultaneous lines.}
         \label{design2}
   \end{figure}

\begin{figure}
   \centering
   \includegraphics[width=\textwidth]{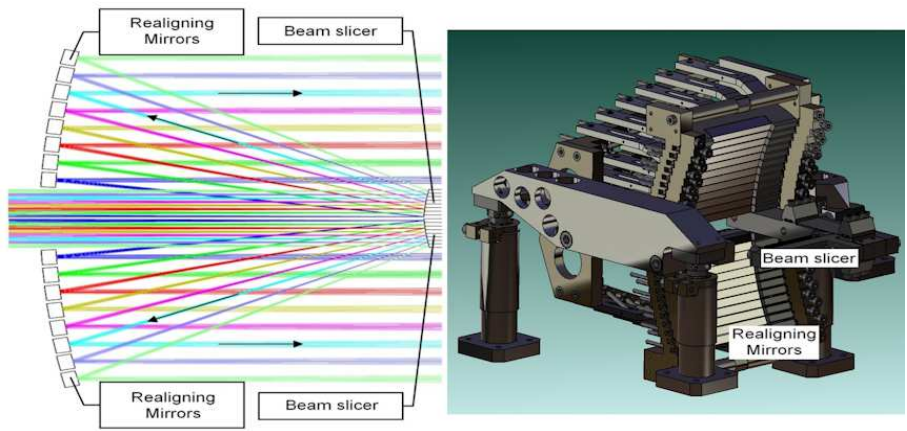}
      \caption{The slicer S is the core of the SLED. It is composed
      of 24 beam-splitting micro-mirrors (0.4 mm step) and 24
      beam-shifting
      mirrors (the figure does not show the full number of mirrors).}
         \label{slicer}
   \end{figure}

The optical path of the SLED spectrograph is given in
Figure~\ref{design2} for the FeX and FeXIV coronal lines. The
letters L and M represent lenses and plane mirrors. Two beams (red
and green) are drawn for each line. The optical parts of the
spectrograph are listed below:

\begin{itemize}
  \item The transfer optics W (the blue box) to the SLED.
  It is detailed in Figure~\ref{design} (left) and contains the two rectangular
  entrance windows (4.4 mm $\times$ 31.0
  mm) of the green (FeXIV) and red (FeX) beams located at the spectrograph focus.
  \item The first pass (light dispersion): W, L1, M1, L2, M2, L3, L4, G (grating in pupil
  plane), L4, L3, M2, L2, M1, L1, M3, S (slicer).
  \item S = 24 channels slicer located in the spectrum (details in Figure~\ref{slicer}).
  It is composed of
  two parts: 24 beam-splitting micro-mirrors and 24 associated beam-shifting
  mirrors.
  \item The second pass (subtractive dispersion): S (slicer), M4, M5, L1, M1, L2, M2,
  L3, L4, G (grating), L4, L3, M2, L2, M1, L1.
  \item The transfer optics to the detector: M6 (two mirrors), L5 (field lens), P (compensating plate), M7 (two mirrors), L6
  (100 mm focal length objective, 0.2 $\times$ magnification, mounted on the
  camera).
  \item The Andor Zyla camera (C, 5.5 Mpixels sCMOS detector, 2560 $\times$ 2160 format, 6.5 micron
  square pixels, USB3, Peltier and air cooled), recording the 24 channel
  spectra-images of the two simultaneous lines.
\end{itemize}

The optical combination of lenses (L1 and L2, L3, L4) acts as a
collimator objective (2.0 m equivalent focal length) with folding
mirrors (M1, M2). The grating (G) is $62^{\circ}$ blazed and 79
grooves/mm ruled; the second pass on the grating subtracts the
dispersion after selection of 24 channels by the slicer S.
Wavelengths of observable lines in the blaze are such that $n \times
\lambda$ = 223531, where $n$ and $\lambda$ are respectively the
order of the spectrum and wavelength (\AA). Broad-band (100 \AA~
FWHM) interference filters (not represented) select the order $n$
(see Table~\ref{capab}).

   \begin{table}
      \caption[]{Capabilities of the SLED.}
         \label{capab}
     $$
         \begin{array}{|c|c|c|c|c|}
            \hline
            \noalign{\smallskip}
            Wavelength & Dispersion    &  Step   & Order & Spectral \\
                (\AA)  & (mm/\AA)      &  (\AA)  &       & resolution~R       \\
            \hline
            \noalign{\smallskip}
            Corona & & & & \\
           6374~FeX & 1.18 & 0.34 & 35 & 19000 \\
           5303~FeXIV & 1.43 & 0.28 & 42 & 19000 \\
            Prominences & & & & \\
           5876~He~D3 & 1.28 & 0.31 & 38 & 19000 \\
           6563~H\alpha & 1.15 & 0.35 & 34 & 19000 \\
            \noalign{\smallskip}
            \hline
          \end{array}
     $$
   \end{table}

\begin{figure}
   \centering
   \includegraphics[width=0.75\textwidth]{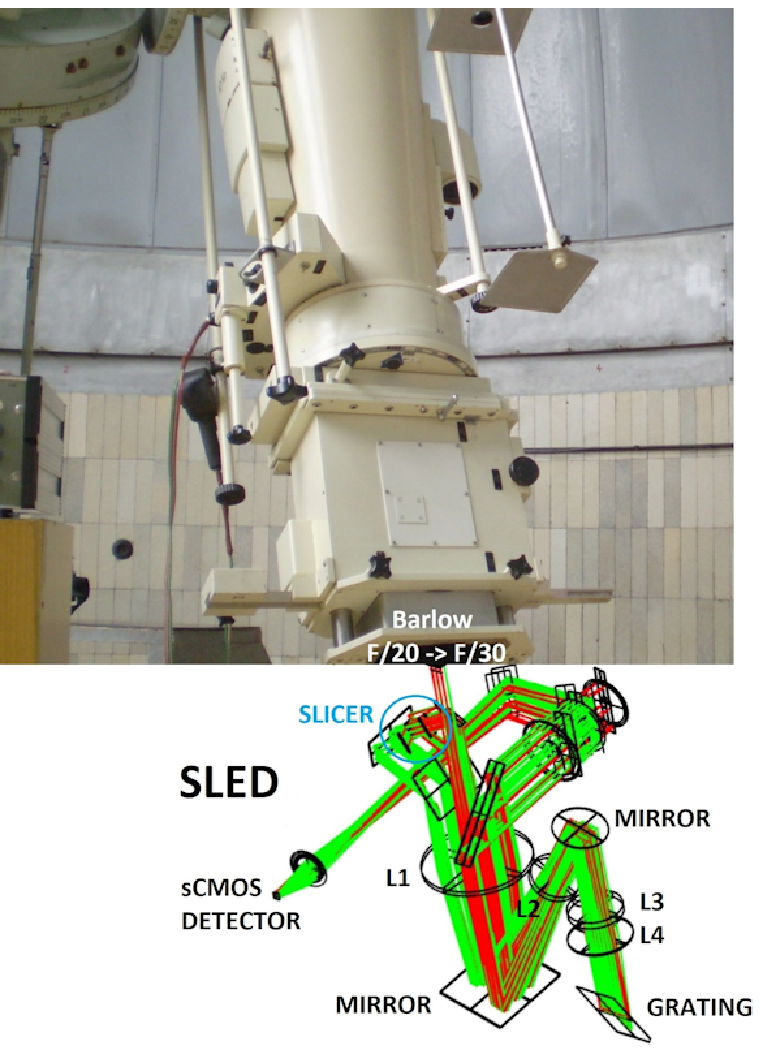}
      \caption{The Zeiss coronagraph at LSO (0.2 m/4.0 m) and the final
      design of the SLED for the green and red lines. The SLED is fed by a 1.5 $\times$
      Barlow lens to work at F/30. It uses extra folding mirrors to
      reduce the volume and optimize the location of the gravity centre.} \label{design3}
   \end{figure}

Figure~\ref{design2} does not exactly represent the final design to
be easily readable. Indeed, the spectrograph will be more compact
using extra folding mirrors that shown in Figure~\ref{design3}.

The slicer \citep{Sayede2014} is composed of two parts: a
beam-splitter of 24 micro-mirrors and a beam-shifter of 24
realigning mirrors, as shown by Figure~\ref{slicer}. The step
between the 24 micro-mirrors is 0.4 mm, providing a spectral
sampling of 0.34 \AA~ and 0.28 \AA, respectively for the FeX 6374
\AA~ and FeXIV 5302 \AA~ coronal lines (Table~\ref{capab}). The
beam-splitter is a solid aluminum alloy component manufactured by
Savimex SAS (France), with its 24 facets polished and aluminum
coated (reflection range 390-900 nm). High quality polishing is
crucial to avoid scattered light. The slicer is located in the
spectrum. SLED's mechanical structure allows the adjustment of the
24 shifting mirrors by two rotations (around directions
perpendicular to the optical axis) and one translation (along the
optical axis). The adjustment's resolution is a few arc seconds for
the tilts and tens of microns for the translation. Each optical path
for the 24 pairs of splitting-shifting mirrors is the same length. A
bench will be used to prepare the slicer, but the final alignment
will be performed in the spectrograph. For each channel, the tilts
allow to centre the pupil on the grating (for good uniformity and
transmittance). The translation adjusts the focus of each channel
and corrects the field curvature.

\begin{figure}
   \centering
   \includegraphics[width=\textwidth]{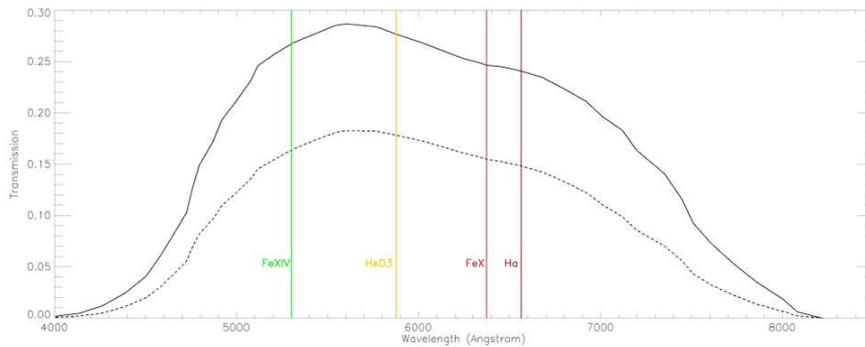}
      \caption{SLED transmission curves as a function of wavelength (\AA). Solid line: optical transmittance;
      dashed line: overall transmittance including the detector.
      Coronal lines (FeX and FeXIV) and prominence lines (He D3 and H$\alpha$) are indicated.}
         \label{trans}
   \end{figure}

Focused spectra-images of both FeX and FeXIV lines, each with 24
channels, are formed simultaneously on the detector. The optical
transmittance of the SLED for these lines is 0.25, as shown by
Figure~\ref{trans}, and is greater than 0.2 over the entire range
5000 - 7000 \AA. There is almost no transmission above 8000 \AA,
which means that IR lines cannot be observed. For instance, the IR
lines (FeXI 7892 \AA~ and FeXIII 10747 \AA) planned to be observed
by the future Visible Emission Line Coronagraph (VELC) onboard the
indian Aditya-L1 mission (\cite{Raj2018}, \cite{Alditya2017}) are
not accessible to the SLED, but coordinated campaigns could be
organized in the FeXIV 5303 \AA~ line with both instruments. The
sCMOS camera has a quantum efficiency greater than 0.5 over the
wavelength range 4250 - 8000 \AA~ and the overall transmittance of
the SLED is reported in Figure~\ref{trans}.

\section{Conclusion} \label{sec:Conclu}

The Solar Line Emission Dopplerometer (SLED) is a state-of-the-art
and innovative instrument for investigating the dynamics of coronal
structures in the forbidden lines of FeX and FeXIV using high
altitude coronagraphs. Dopplergrams (150$''$ $\times$ 1000$''$ FOV,
2.1$''$ pixel sampling), providing the line-of-sight velocity, will
be produced at high cadence (1 Hz). This original feature will allow
us to observe plasma motions during fast-evolving events such as
flares or CMEs driving space weather phenomena, and also
high-frequency oscillations and waves implied in coronal heating.
Full line profiles (0.28 \AA~ resolution) will be available for all
pixels of the FOV. The SLED will complement SDO/AIA (EUV intensities
of hot lines) and ground-based coronal imagers such as SECIS. It
uses an imaging spectroscopy technique which is faster than most
tunable filters and narrow slit spectrographs. The SLED will
supplement spectroscopic instruments operating mainly in IR lines,
such as CoMP or the future VELC instrument onboard Aditya-L1
\citep{Raj2018}. The Multi-slit Solar Explorer (MUSE) project
\citep{Depontieu2020} is another approach for EUV spectroscopy in
space. SLED will follow the coronal activity of cycle 25 at
Lomnick\'{y} \v{S}t\'{\i}t Observatory, but will also be tested at
Bia{\l}k\'{o}w to study the dynamics of prominences. As it is a
portable instrument, it could also be deployed to observe total
solar eclipses, such as the event of 8 April 2024 visible from North
America.

\begin{acknowledgements}

The authors thank the referee for helpful comments and suggestions.
We are grateful for financial support to the Institut National des
Sciences de l'Univers (INSU/CNRS), the University of Wroc{\l}aw, the
UK Science and Technology Facilities Council (STFC), the Leverhulme
Trust via grant RPG-2019-371, and Queen's University Belfast. J.R.
acknowledges support by the Science Grant Agency project VEGA 2/0048/20 (Slovakia). P. Rudawy were supported by the National Science Centre in Poland, under grant No. UMO-2015/17/B/ST9/02073.
\\

\end{acknowledgements}

\bibliographystyle{spbasic}      

\bibliography{papier}

\begin{appendix} 

\section{Online material}

MPEG4 movie showing a simulation of the 24 channel spectral images
of the SLED for the coronal green line (FeXIV 5303 \AA, top) and red
line (FeX 6374 \AA, middle). Three rectangular structures have been
chosen (centre, left and right of the FOV) with LOS velocities
continuously varying from -100 km s$^{-1}$ to +100 km s$^{-1}$
(intensity I and velocity V at the right of the display, intensity
in green with blueshift and redshift).

The bottom pannel shows the corresponding line profiles (left: green
line; right: red line), for structures located in the centre and
both sides of the FOV. The wavelength (\AA) is centred. The spectral
range (6.5 and 8.0 \AA~ respectively for the green and red line) is
constant across the FOV but shifts according to the x-position.
Velocities up to $\pm$75 km s$^{-1}$ can be measured in any point of
the FOV, and much more near the centre.

\end{appendix}

\end{document}